\documentclass[unnumsec,webpdf,modern,large]{oup-authoring-template}

\graphicspath{{Fig/}}

\theoremstyle{thmstyleone}%
\theoremstyle{thmstyletwo}%
\theoremstyle{thmstylethree}%

\begin{document}

\journaltitle{Journal of the American Medical Informatics Association}
\DOI{DOI HERE}
\copyrightyear{2024}
\pubyear{2014}
\access{Advance Access Publication Date: Day Month Year}
\appnotes{Paper}

\firstpage{1}


\title[AI and ECG]
{Contrasting Attitudes Towards Current and Future AI Applications for Computerised Interpretation of ECG: A Clinical Stakeholder Interview Study}

 \author[1,2,$\ast$]{Lukas Hughes-Noehrer, PhD \ORCID{0000-0002-9167-0397}}
 \author[1]{Leda Channer, PhD}
 \author[1]{Gabriel Strain, MRes}
 \author[2]{Gregory Yates, MBBS}
 \author[2,3]{Richard Body, PhD}
 \author[1]{Caroline Jay, PhD}

 \authormark{Hughes-Noehrer et al.}

 \address[1]{\orgdiv{Department of Computer Science}, \orgname{The University of Manchester}, \orgaddress{\street{Oxford Road}, \postcode{M13 9PL}, \country{United Kingdom}}}
 \address[2]{\orgname{Manchester University NHS Foundation Trust}, \orgaddress{\street{Oxford Road}, \postcode{M13 9WL}, \country{United Kingdom}}}
 \address[3]{\orgdiv{Division of Cardiovascular Sciences}, \orgname{The University of Manchester}, \orgaddress{\street{Oxford Road}, \postcode{M13 9PL}, \country{United Kingdom}}}

 \corresp[$\ast$]{Corresponding author: \href{lukas.noehrer@manchester.ac.uk}{lukas.noehrer@manchester.ac.uk}}

\received{Date}{0}{Year}
\revised{Date}{0}{Year}
\accepted{Date}{0}{Year}



\abstract{\textbf{Objectives:} To investigate clinicians’ attitudes towards current automated interpretation of ECG and novel AI technologies and their perception of computer-assisted interpretation.\\
\textbf{Materials and Methods:} We conducted a series of interviews with clinicians in the UK. Our study: (i) explores the potential for AI, specifically future ‘human-like’ computing approaches, to facilitate ECG interpretation and support clinical decision making, and (ii) elicits their opinions about the importance of explainability and trustworthiness of AI algorithms.\\
\textbf{Results:} We performed inductive thematic analysis on interview transcriptions from 23 clinicians and identified the following themes: (i) a lack of trust in current systems, (ii) positive attitudes towards future AI applications and requirements for these, (iii) the relationship between the accuracy and explainability of algorithms, and (iv) opinions on education, possible deskilling, and the impact of AI on clinical competencies.\\
\textbf{Discussion:} Clinicians do not trust current computerised methods, but welcome future `AI' technologies. Where clinicians trust future AI interpretation to be accurate, they are less concerned that it is explainable. They also preferred ECG interpretation that demonstrated the results of the algorithm visually. Whilst clinicians do not fear job losses, they are concerned about deskilling and the need to educate the workforce to use AI responsibly.\\
\textbf{Conclusion:} 
Clinicians are positive about the future application of AI in clinical decision-making. Accuracy is a key factor of uptake and visualisations are preferred over current computerised methods. This is viewed as a potential means of training and upskilling, in contrast to the deskilling that automation might be perceived to bring.}


\maketitle

\section{INTRODUCTION}

Heart and circulatory diseases cause approximately 480 deaths per day in the United Kingdom alone \cite{the_british_heart_foundation_uk_nodate}. Electrocardiograms (ECG) support the diagnosis of cardiac abnormalities based on the heart’s electrical activity. Using algorithms to assist the interpretation of ECGs dates back to the 1950s \cite{macfarlane_automated_2021}, and they are now a staple functionality of ECG monitoring.

Computerised interpretation of ECG (CIE) is designed to support clinical decision-making, reduce diagnostic errors, and improve patient outcomes. Whilst the performance and accuracy of these algorithms has improved considerably over the years, studies have shown that contemporary methods are still error prone, requiring oversight from a clinician to avoid misdiagnosis \cite{smulyan_computerized_2019, schlapfer_computer-interpreted_2017}.
With the establishment of large waveform datasets and advances in Artificial Intelligence (AI) and Machine Learning (ML) techniques, methods to process and interpret ECGs now range from rule-based systems to deep learning approaches \cite{gupta_computer-interpreted_2024, gill_artificial_2023, lyon_computational_2018}.

Whilst AI applications have been advancing the field of CIE \cite{yao_ecg_2020, noseworthy_artificial_2022, martinez-selles_current_2023}, they do not necessarily meet clinical needs, impeding wider adoption \cite{khurshid_clinical_2023, siontis_artificial_2021, elul_meeting_2021}. Reasons for non-adoption of AI applications in clinical practice include issues of systemic bias, open questions around liability, cyber security concerns, unresolved technical challenges, risks of AI exacerbating health inequities, and ethical and regulatory challenges \cite{chung_clinical_2022,rajpurkar_ai_2022}. Furthermore, there is a lack of real-world evidence for service providers, such as the UK’s National Health Service (NHS) \cite{lip_adoption_2024, castagno_perceptions_2020}, and more broadly, there are insufficient governmental policies and economic incentives \cite{panch_inconvenient_2019} to drive wider uptake of AI applications in the health domain.

A key issue is a lack of trust in AI systems \cite{burgess_healthcare_2023, asan_artificial_2020}. A mixed-methods study investigating AI in cardiovascular medicine concluded that clinicians generally feel positive towards AI and its future potential, but also highlighted several barriers, including limited trust in its output, high costs, and insufficient usability \cite{schepart_artificial_2023}. An interview study conducted with Australian emergency physicians concluded that clinicians generally believe that AI will have an impact on their field within the next decade, but further work is necessary to improve its acceptance \cite{stewart_attitudes_2024}. Explainable algorithms based on human-like computing have shown efficacy in making complex ECG data easier to interpret for lay people and medical professionals alike \cite{alahmadi_explainable_2021}. A study investigating explainable ML models indicated clinicians felt generally positive towards AI, but noted issues such as confirmation bias and the cognitive complexity of reasoning with current explainable algorithms \cite{wysocki_assessing_2023}.

Successful implementation of ML-driven Decision-Support Systems requires alignment of the AI application with clinicians' expectations by bridging the `sociotechnical gap' \cite{matthiesen_clinician_2021}. Expectations of developers and clinicians often differ in terms of their understanding of concepts such as explainability, and common frameworks could help to overcome barriers and find agreement between stakeholders \cite{bienefeld_solving_2023, henry_humanmachine_2022}. A study investigating radiologists' and computer scientists' views on AI highlighted significant differences between these groups, with the latter generally predicting faster implementation and a greater impact \cite{eltorai_thoracic_2020}.
AI applications must be trusted by all stakeholders to ensure an inclusive, ethical, and sustained implementation \cite{goldberg_no_2024}.

Here, we present a qualitative interview study conducted in the course of our clinical stakeholder engagement to develop a novel `human-like' AI algorithm for ECG interpretation. Our algorithm is based on cognitive fit theory \cite{shaft_role_2006} and provides a novel visualisation of ST-elevation myocardial infarction (STEMI) to enhance its recognition by healthcare professionals and lay people. 

We interviewed 23 clinicians across three specialties (emergency medicine; anaesthesia and critical care; and cardiology), covering all training stages of the UK medical system. We also included general practitioners (GP) and junior doctors in their foundation training. This paper examines the attitudes of clinicians towards current computer-automated ECG interpretation and also novel AI algorithms that could be used to support future diagnosis. Our research contributes to the understanding of the clinical use of these technologies and how they are perceived by those clinicians who use ECGs on a daily basis. 
The results demonstrate an interesting paradox. Whilst current automated diagnoses are often ignored due to concerns about their accuracy and trustworthiness, and clinicians are concerned that the use of them may lead to deskilling and a negative impact on training, they are positive about future AI applications. So long as algorithms are accurate, clinicians are unconcerned with the explainability of future AI systems. We discuss the relationship between accuracy and explainability, the possible role of AI in education, and the potential impact of future applications on the clinical workforce.


\section{METHODS}

This paper considers the experiences of clinicians with automated ECG interpretation and their attitudes towards novel AI techniques that could be used to facilitate interpretation of ECGs in the future. Participants were sampled across three specialties, GPs, and junior doctors. All clinicians were based in the UK at the time of interviewing. These specialties were selected after a series of Responsible Research and Innovation (RRI) workshops, which identified these groups as regularly interpreting ECGs. The final selection was agreed by the research team formed of clinicians, Human-Computer Interaction researchers, and an RRI specialist. The design of the study was informed by the Standards for Reporting Qualitative Research (SRQR) to ensure rigorous data collection and reporting \cite{obrien_standards_2014}. Based on the Ethics Decision Tool at [omitted] and the NHS Health Research Authority Ethics Toolkit \cite{nhs_ethics}, the study was exempt from ethical review as we solely interviewed professionals (i) strictly within their professional remit, and (ii) did not collect any person identifiable information apart from those on the exempt list, i.e. names, professional roles, signed consent, and audio recordings. An outcome letter is available at Figshare [omitted].

The interviews were undertaken between December 2023 and April 2024 either via the video conferencing software Microsoft Teams \cite{ms-teams} (video function disabled) or in-person at [omitted] and [omitted]. Participants were convenience sampled \cite{mason_qualitative_2018} on the basis that they are currently employed clinicians with “experiential relevance” \cite[p. 124]{rudestam_surviving_2015} in terms of ECG usage and clinical knowledge. Although most participants provided written consent for attribution, as we aimed to generate data that establishes a representational account of the interviewees’ professional roles, we refrain from including the names of participants. All participants received a Participant Information Sheet (PIS) prior to the interview and provided informed consent. Participants were further given the option of withdrawing data up to 14 days after the date of participation. People received a shopping voucher worth Fifty Pound Sterling (GBP) as a token of appreciation for their time investment.

\begin{figure*}[h]%
\centering
\includegraphics[width=\textwidth]{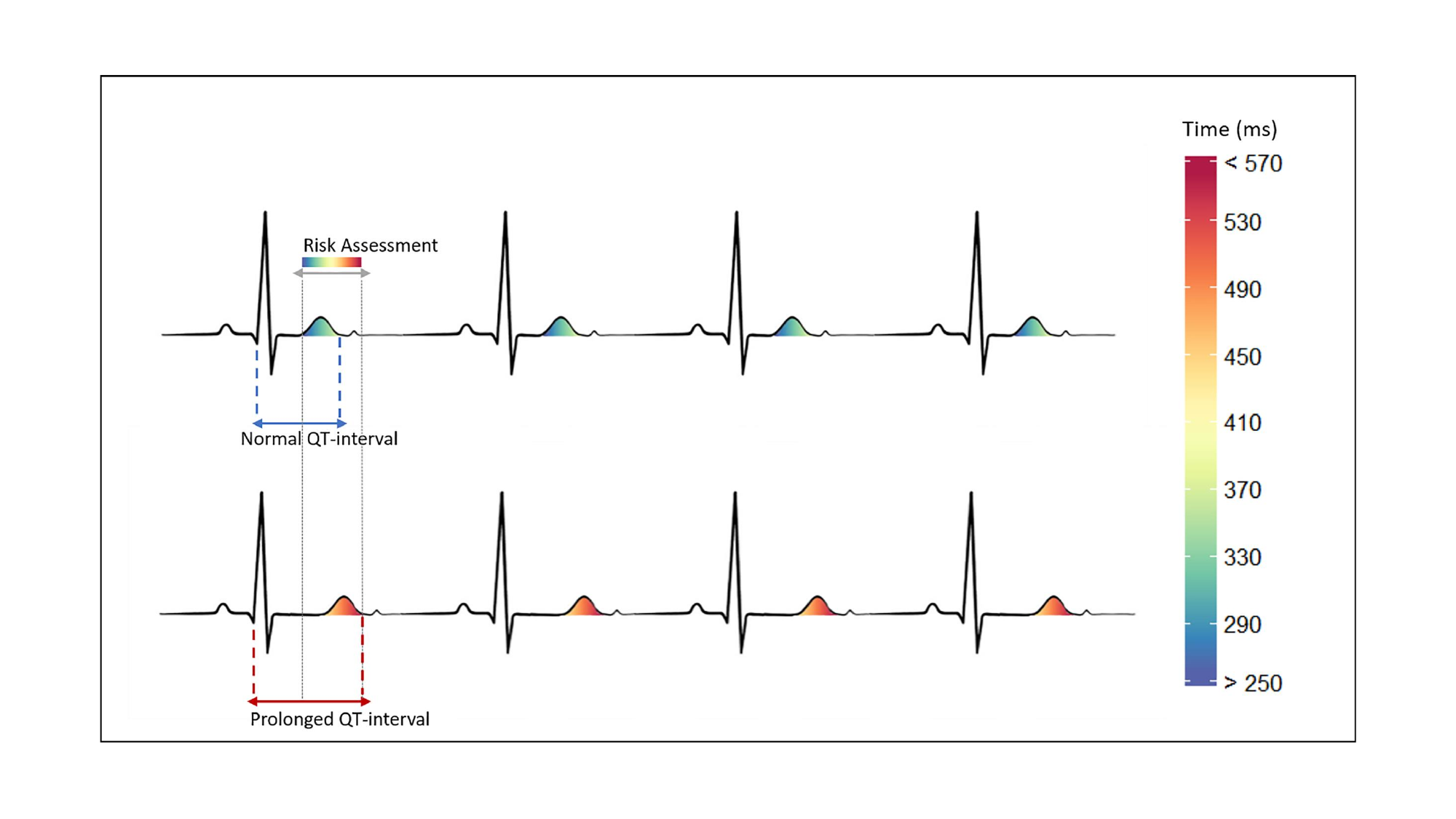}
\caption{Example of a visualisation using our explainable algorithm shown to participants during interviews. Cooler colours indicate a normal QT interval (upper ECG) whilst warmer colours indicate a prolonged QT interval (lower ECG).}\label{fig1}
\end{figure*}

Interviews followed an interview guide \cite{lindlof_qualitative_2017} based on generative questions \cite{rubin_qualitative_2005} to encourage extensive replies in an open format. The topic guide was split into two parts: a general part, which asked clinicians about how often they look at ECGs in their clinical practice, the methodology they use to interpret them, and the pitfalls and strengths of current systems. Part two showed clinicians visualisations (see Figure \ref{fig1}) based on explainable `human-like' algorithms, to elicit their attitudes towards AI technologies and explainability of algorithms. Interviewees were also asked about how our visualisations compare to current CIE. All materials are openly available on Figshare; please refer to the Data Availability Statement.

All interviews were transcribed and cross-checked to ensure reliability and validity \cite{kvale_doing_2007}. Analysis was performed using inductive thematic analysis \cite{braun_using_2006} and the software NVivo 14 \cite{nvivo} to analyse, code, and re-code the interview data, matching sentences to the primary codes in the interview. After generation of primary codes (X.X.), authors discussed and finalised the codes (X.X., X.X. and X.X.) and attributed them to emerging themes. These themes were then re-evaluated and agreed between authors (X.X., X.X., X.X. and X.X.). Transcripts of the interviews are openly accessible and were deposited on Figshare with consent from participants, as described in the data availability statement.

\section{RESULTS}

We interviewed 23 clinicians who, at time of interviews, were actively practising in the NHS. The split of specialties and training levels of interviewees can be found in Table \ref{tab1}. None of the interviewed participants withdrew any of the recorded data and we therefore included all 23 transcripts in the analysis. Interviews had a median length of 21 minutes (Q1: 17 min; Q3: 25 mins; IQR: 8 min). We report the results below under subheadings of identified themes.

\begin{table}[!t]
\caption{Participant demographics split by specialty and training stage\label{tab1}}%
\begin{tabular*}{\columnwidth}{@{\extracolsep\fill}llll@{\extracolsep\fill}}
\toprule
Participant Number (P) & Specialty  & Training Stage \\
\midrule
01    & An\&CC       & ST1-3  \\
02    & An\&CC       & ST1-3  \\
03    & An\&CC       & ST4+  \\
04    & An\&CC       & Consultant  \\
05    & An\&CC       & Consultant  \\
06    & Cardiology   & ST1-3  \\
07    & Cardiology   & ST4+  \\
08    & Cardiology   & ST4+  \\
09    & Cardiology   & ST4+  \\
10    & Cardiology   & ST4+  \\
11    & EM           & ST1-3  \\
12    & EM           & ST1-3  \\
13    & EM           & ST1-3 \\
14    & EM           & ST4+  \\
15    & EM           & ST4+  \\
16    & EM           & ST4+  \\
17    & EM           & SAS/Staff Grade/Fellow  \\
18    & EM           & Consultant  \\
19    & GP           & 1-5 Years  \\
20    & GP           & 1-5 Years  \\
21    & GP           & \textgreater 15 Years  \\
22    & JD           & F1/F2  \\
23    & JD           & F1/F2  \\
\botrule
\end{tabular*}
\begin{tablenotes}%
\item Notes: Anaesthesia and Critical Care (An\&CC); Emergency Medicine (EM); General Practitioner (GP); Junior Doctor (JD). A comprehensive list outlining UK medical doctor titles and training stages can be found here \cite{bma_training}.
\end{tablenotes}
\end{table}

\section{Lack of trust in current systems}

Clinicians have serious concerns around the usage of current CIE for clinical decision making. The majority of interviewees do not trust current systems for multiple reasons.

Most interviewees reported negative experiences using CIE in clinical practice, with participants stating that they are \textit{``wildly inaccurate"} (P04), \textit{``dangerous"} (P15), and \textit{``very unhelpful"} (P09). Computerised interpretation was also seen as lacking context to make informed decisions, leading to a lot of misinterpretations. Clinicians have also been told at different stages throughout their career to ignore them. One clinician reported that they were first told in medical school not to rely on automated interpretations of ECGs and that they should calculate everything themselves (P22). Other interviewees stated that \textit{``probably the first thing you get taught when you interpret an ECG is to ignore the automated rule"} (P18) and that they were \textit{``advised to not rely on them"} (P17).

Interviewees also expressed that the CIE sometimes caused unease when looking at an ECG. This is due to automated read-outs suggesting diagnoses that are not present, causing added stress as the technology indicates issues that do not align with the clinician’s own interpretation. P22 recalled a recent situation:\textit{``There was a patient who had an ECG that I reviewed, and it [the ECG monitor] said `STEMI' at the top, and for the life of me, I could not see the STEMI. And it was, it was terrible. I was terrified."}

A few clinicians stated that they do use current CIE, but they exercise caution and mainly compare their own findings to the computer rather than relying on them to inform their diagnosis from the start.

\section{Future application of AI}

Despite the shortfalls of current CIE, clinicians were broadly positive about novel AI technologies, appearing to see them as qualitatively different to current methods. All interviewees agreed that digital technologies will positively impact the way we record and interpret ECGs in the future. A qualification to this optimism was that deployment of AI technologies to facilitate interpretation of ECGs should be context-dependent, and it should be possible to customise applications.

Concerns raised about AI included not having access to the original data once it has been computationally processed, which could introduce the risk of missing diagnostically valuable data: \textit{``[...] if it was drawing attention to one thing, you'd be necessarily neglecting another area, or it might cull your own interpretation"} (P04). To counteract this, interviewees would appreciate options to add AI capabilities on-demand, e.g., via filters or button settings. The visualisations indicating how the algorithm was using the data to make its interpretation were viewed positively, as they would allow the clinician to focus on what the AI thought was important.

When asked about their preference for AI highlighting abnormalities on an ECG (e.g., ST-elevation) vs. specific conditions (such as STEMI), clinicians' opinions were divided. Some felt unease at the idea of AI providing a diagnosis. P17 remarked that AI models -- as they are currently developed -- lack the important contextual information necessary to make a clinically valid diagnosis, as the data they use is limited to the ECG signals. Drawing attention to anomalies in the signal rather than a full diagnosis leaves more room for the clinician's own interpretation and decision-making, within the wider patient context. 

If reliable and accurate, however, a condition-based approach could be a valuable tool for training and developing ECG interpretation skills. In these circumstances, clinicians preferred to have visualisations that demonstrate which areas have been considered by the AI, as it was thought to increase explainability and clinicians' learning.

Clinicians want applications that fit seamlessly into their practice, rather than disrupting it by adding additional uncertainty. P16, who, when acting as Emergency Physician in Charge, signs off ECGs done in the department, hoped that interpretation technology would be able to bring \textit{``some relief"}. This was also the view of P01, another clinician who has to sign off ECGs and supervise more junior colleagues, who thought that AI could bring \textit{``another bit of reassurance"}.

\section{Accuracy and explainability}

When asked about the explainability of AI, the majority of clinicians stated they did not necessarily have a preference for understanding how an algorithm reached its decision, as long as high accuracy was guaranteed.

In spite of positive views of the visual representations of AI decision-making, there was a view that \textit{``moving towards non-explained AI is probably the way things are gonna go"} (P19) and clinicians are prepared to trust future AI as long as the system is accurate and reliable enough to include its outputs into their clinical decision-making. Another interviewee remarked that current systems are already perceived as “black box” algorithms (P01), as contemporary clinical ECG applications do not tell clinicians about how they arrive at conclusions. A similar comparison was drawn by other clinicians, who described clinical decision-making itself as a black box, as they felt that a lot of experience of interpreting ECGs leads to \textit{``a bit of instinct and gut to rely on"} (P12).

Clinicians also noted that specifically in emergency situations there is no time to reason about a computer's output, as the care of patients has the highest priority. This was summed up by P16 who stated: \textit{``In clinical practice, I actually want to know less, because we are overwhelmed with so many things."}

A few clinicians thought that explainable algorithms would be valuable. Reasons for this were enhanced trust, and knowledge about an algorithm's weaknesses, but also from a point of care to show competency to patients and to \textit{``give them some comfort to know that we understand the machines we're dealing with"} (P23). Some interviewees also saw value in explainable algorithms as they have some personal interest or experience with machine learning, and knowing what was going on behind the scenes would be driven by personal curiosity.

\section{Patients and AI}

Interviewees were ambivalent when asked if they think their patients would prefer explainable AI or not. Most clinicians stated that patients generally neither see their ECG nor are they involved in the process of interpreting them, highlighting that patients are often not involved in the clinical decision-making progress enough to understand it. Other interviewees saw a responsibility toward patients to educate them about the use of AI and to inform them that algorithms formed part of their decision-making process. P20 remarked that some patients embrace new technologies and see benefits in approaches such as deep learning, which might become even more prevalent in digital-native generations.

There was also a notion that too much information can cause unwanted outcomes: \textit{``I like to educate my patients and explain -- `you know, this is what this means' -- but I find actually nine times out of ten, it just gives them more anxiety when you explain things to them"} (P12).

Clinicians overall agreed that patients want accurate systems that clinicians can trust and to know that clinicians are responsible for patient outcomes, not a computer.

\section{Education, deskilling, and clinical competencies}

Whilst none of the clinicians raised concerns about AI making their roles redundant, they highlighted possible ways in which AI might impact medical education and clinical practice.

There was a view that AI might contribute to a potential loss of skills in future generations of doctors. Clinicians mainly feared that if an AI system just spits out diagnoses, this could mean doctors \textit{``stop practising interpretation"} (P04). P22 highlighted that \textit{``there were a lot of worries about juniors who, but almost kind of like born into that system, that there would be certain skills that would be lost."} However, others saw potential in AI-aided applications for educational purposes, as algorithms could be used to specifically visualise or explain ECGs to medical students or doctors in early career stages. A condition-based approach was seen as a valuable tool for training and developing ECG interpretation skills. Visualisations were viewed as useful for telling clinicians which areas have been considered by the AI, increasing explainability and providing a shared representation between human and machine:  \textit{``So actually you almost think the colour...you know, the fact that part of it is highlighted, shows to me that the system or the black box has spotted that bit and can colour it in, and therefore that automatically adds weight to whatever conclusion it draws."} (P01)

Participants raised concerns about over-reliance on computerised interpretation in clinical practice. Interviewees feared that AI may deskill clinicians, making them \textit{``lazy"} (P04) when a computer just gives them an answer, but they also remarked that losing skills has been happening for a while. Some clinicians noted that AI is already advancing into several fields and that the medical profession just has to accept that these systems will arrive in the future. For example, P04 stated: \textit{``I'd say yeah, fine, takes workload off us, and we lose skills, but I mean, that's the nature of medicine, we always lose skills as technology develops."} The responsibility of the clinician for a decision remained paramount, however. The majority of interviewees said that although AI might help to interpret ECGs, it is important to safeguard clinical competencies and skills.


\section{DISCUSSION}

The clinicians we interviewed do not trust current CIE as a basis for clinical decisions as these systems are deemed unreliable and often over-sensitive. Clinicians also reported adverse effects of computerised interpretation on clinical decision-making, leaving young doctors or those with less training worrying about the accuracy of their own diagnoses, as they did not feel they had enough experience of interpreting ECGs to confidently ignore CIE. 
The literature we reviewed confirmed that current CIE systems are often not used and that there are issues with their reliability \cite{smulyan_computerized_2019, schlapfer_computer-interpreted_2017}. 

In contrast, there was a view that novel AI applications will outperform current commercial products \cite{van_de_leur_automatic_2024, herman_validation_2024}. Furthermore, our findings indicate that clinicians do not associate current automated methods with AI-driven applications, which could be a factor in AI uptake and acceptance.

AI applications should fit seamlessly into clinical practice and gain clinicians' trust  by delivering accurate results requiring low cognitive effort to understand. There is no one-size-fits-all approach as clinicians have varying concerns and demands that will have to be met to ensure usage. The ability to interact with the system and to adapt it to individual requirements would be of value.

As long as algorithms were accurate, explainability was not viewed as important by many clinicians. This finding is congruent with research conducted with pathologists who do not seem to worry about explainability as long as accuracy and usefulness can be guaranteed \cite{king_how_2023}. In fast-paced, clinical environments such as emergency settings, explainability may play a lesser role as rapid decision-making is required and information overload can hinder this.


Participants raised concerns about technology deskilling clinicians. Although they did not all require explanations for a diagnosis, they also indicated they would prefer to have a visualisation of how a decision has been made. 
Visualisations that indicate abnormalities on ECGs were viewed as useful for avoiding deskilling and supporting clinical decision-making. 

As clinicians appear prepared to increasingly rely on technology, we suggest that digital technology education should provide a foundation in how it works in practice, including: (i) what AI actually is, (ii) the tasks it can perform, and (iii) how models make decisions. This will help prepare the workforce of the future to make the best use of upcoming AI technologies in practice.

\section{CONCLUSION}

Our research finds that clinicians do not trust current automated ECG interpretation but have a positive attitude toward novel AI technologies. Explainability is valued but not essential, as long as the system is accurate. In order to function well, as system must fit into clinical practice and avoid imposing an additional cognitive burden. 
Early education, starting in medical school, will prepare future doctors for ethical and informed use of future technologies and support them in making AI-informed decisions. 



\section{Limitations}

Our study is limited to the views of 23 clinicians. As this is a limited sample that comes from NHS trusts across England, we do not aim to present results that generalise across a wider population, but rather aim to contribute further understanding and context around AI from the perspective of three specialties, GPs, and junior doctors. Although some of the clinicians were not trained in the UK, this study presents a UK-centric position of English-speaking clinicians. We acknowledge that future work should focus on the incorporation of further specialties, such as emergency medical services (EMS; paramedics), PPIE stakeholders, and inclusion of a broader variety of demographics.

\section{Competing interests}
The authors have no competing interests to declare.

\section{Author contributions statement}


\section{Acknowledgments}

\section{Data availability statement}

The data underlying this article is available in Figshare [omitted]. The transcript data is openly available and reusable under CC BY 4.0. The topic guide, PIS, an exemplar consent form, and the visualisations shown to participants are also included. 
For context, this paper presents a specific focus on clinicians' perception and attitudes of automated ECG interpretation and AI, however, we collected further information, such as how clinicians use and interpret ECGs, how often they do so, their methods used to interpret ECGs, and which pathologies are particularly difficult to spot.

\bibliographystyle{unsrt}
\bibliography{reference}

\end{document}